# Fusing location and text features for sentiment classification

1st Wei Lun Lim
Faculty of Computing & Informatics
Multimedia University
Cyberjaya, Malaysia
0000-0002-0301-2846

2nd Chiung Ching Ho
Faculty of Computing & Informatics
Multimedia University
Cyberjaya, Malaysia
0000-0001-5098-8131

3rd Choo-Yee Ting
Faculty of Computing & Informatics
Multimedia University
Cyberjaya, Malaysia
0000-0001-5667-2816

*Abstract*—Geo-tagged Twitter data has been used recently to infer insights on the human aspects of social media. Insights related to demographics, spatial distribution of cultural activities, space-time travel trajectories for humans as well as happiness has been mined from geo-tagged twitter data in recent studies. To date, not much study has been done on the impact of the geo-location features of a Tweet on its sentiment. This observation has inspired us to propose the usage of geo-location features as a method to perform sentiment classification. In this method, the sentiment classification of geo-tagged tweets is performed by concatenating geo-location features and one-hot encoded word vectors as inputs for convolutional neural networks (CNN) and long short-term memory (LSTM) networks. The addition of language-independent features in the form of geo-location features has helped to enrich the tweet representation in order to combat the sparse nature of short tweet message. The results achieved has demonstrated that concatenating geo-location features to one-hot encoded word vectors can achieve higher accuracy as compared to the usage of word vectors alone for the purpose of sentiment classification.

## I. INTRODUCTION

The continual improvements in communication infrastructure and increased access to personal mobile devices has led to the dramatic increase in social media content. In some instances, the social media data is encoded with spatial temporal features including global positioning system (GPS) coordinates and names and types of nearby locations – which results in what is known as geo-social media[1].

One important aspect of geo-social media research is to examine the complex relationship between the users of geo-social media and their environment. Understanding this relationship can yield insights concerning the users of geo-social media, as well as the environment in which the geo-social media content is generated.

Geo-social media data have been used to infer observable aspects of geo-social media users, especially in the movement and location of users. The home location of geo-social media users can be inferred by using the location shared in tweets, with an accuracy rate of 80% from a sample rate of as low as 1.5 tweets per day [2]. Another study focused on the movement of users, which was derived with 95% accuracy rate using data collected over a 12 month period [3]. The success in inferring the location and movement of users via their geo-social media content has led to increased applications in the tourism sector, with studies being performed on itinerary [4] and location recommendation [5], as well as the tracking of number of visitations to tourist sites[6] and tourist-resident segmentation [7].

Non-observable aspects of a geo-social media user have also been studied, with a focus on the user's motivation for using geo-social media, as well as for localized opinion mining and reactions towards an event in the users' locality. In terms of motivation, geo-social media users often participated in a local social network in order to foster social integration and build ties, in the same way as a local newsletter [8]. Aside from community engagement, localized opinion mining can be performed using geo-social media, and have been used to elicit a response for participatory urban planning [9]. Opinion mining is often linked to responses towards an event, and collective reactions to an event has been studied [10] to infer a collective reaction across multiple geo-social media users.

The literature review performed during our study have clearly established that there is a relationship between geo-social media users and their environment. Spatial temporal data contributed by geo-social media have resulted in a "quantified environment", whereby an environment is measurable across time and space with regards to the attributes of its users, both intrinsic and extrinsic.

The establishment of this relationship has led to the research question of whether the relationship between geo-social media users and their environment is one-way. There has been studies which reported a strong relationship between one's physical environment and one's mood and cognitive performance [11], as well as one's tendency to indulge in impulse buying[12]. As this area of research is relatively unexplored, we were thus motivated to study whether the environment impacted the geo-social media user's sentiment, in as much as the social-media user impacts the environment.

The goal of this study is to examine the impact of geo-location features from an environment on the sentiment expressed by users of geo-social media in the same environment.

In our study, we focus on geo-tagged tweets, with the objective to determine whether the concatenation of geo-location features to word vectors can result in increased accuracy of the sentiment of geo-tagged tweets. We examined a geo-tagged twitter dataset which has been tagged with its sentiment value using our approach.

## II. PROPOSED APPROACH

The challenge of inferencing the sentiment of geo-tagged tweets is many-fold. Firstly geo-tagged tweets by their nature of being limited to 140 characters suffers from sparsity of data [13]. Secondly, the dependence of sentiments on phrases [14] requires that sequence information be preserved in order to achieve higher accuracy, which hampers bag-of-words approaches for tweet representation. Thirdly, the limited number of archival geo-tagged tweet datasets results in a relatively small pool of candidate data to be used as training for any word embedding model.

In this context, the proposed approach used in our study is inspired by previous solutions to the previously stated challenges. For the first challenge, concatenated word embeddings have been used to the improve performance for a word analogy task [15] in order to address the sparsity of the tweet representation. For the second challenge, the usage of a long short term memory (LSTM) architecture [16] for the purpose of maintaining sequential information have increased the accuracy of a tweet polarity task. For the final challenge, we have leveraged on the idea of using pre-trained word embeddings which was used to achieve good results on a sentiment classification task [17] which suffered from limited opportunities for training of a custom word embedding model [18].

### A. Problem definition

In this study, the dataset used was modified from the Geo-tagged Microblog Corpus [19] which contains geotagged tweets. The original dataset contained 377616 tweets, from which a subset of 10000 tweets was created via random sampling.

### B. Text preprocessing and sentiment labelling

In this study, the sentiment of each tweet was determined based on purely text content. Non-text information which includes Unicode Strings, Numbers, Website links and URLs, Retweet and Mentions symbols, Hashtag symbols, Punctuation, as well as White Spaces was removed.

Following the removal of non-text information, the sentiment score for each tweet was using pattern matching of phrases implemented in the TextBlob [20] library. Each tweet was then classified to either a positive or negative class using the algorithm shown in Fig.1.

**Algorithm** : Classification of tweets to positive or negative sentiment

**Data** : Tweets
**Results**: categorize tweets sentiment to 0 or 1 based on sentiment score

```
while not at end of this document do
read current
        if score > 0 then
                label sentiment score as 1;
        else label sentiment score as 0;
```

Fig. 1. The algorithm used to label each tweet as exhibiting positive (sentiment score =1) or negative sentiment (sentiment score = 0) is shown in pseudocode

### C. Geo-location features elicitation

Each tweet is processed to elicit its geo-location features. This is achieved by using the Geonames Find Nearby Web Services [21] and the Google Places Nearby Search [22] API to extract the categories of locations that is nearby the location of each tweet, as each tweet has its GPS coordinates. The nearby locations were extracted within a 300 meters radius of each tweet. The categories of nearby locations information are then vectorized before being appended to each tweet. The extraction of the categories of nearby location can be time consuming, as the number of free API calls allowed per day are limited.

### D. Text and geo-location feature concatenation

All tweets were padded to the length of 25, as this is the maximum length of the tweets after the tweets were integer encoded during vectorization. The vectorized tweets are then concatenated with the vectorized nearest location categories, resulting in a feature vector of length of 76 and 125, depending on whether the nearby locations were extracted using the Geonames and Google Places web services respectively. The length of the feature vector reflects the total categories returned by the Geonames (51 categories) and Google Places (100 categories) web services. In this study, we used two ways to generate the vectorized categories of nearby locations, namely one-hot encoding and count. The one-hot nearby location categories encoding transformed the nearby location categories into a binary vector consisting of the occurrences of said location category. Using the count method, the nearby location categories were converted into a numeric vector of occurrences, which has the value ranging from zero(non-occurrence) to N (where N is the number of occurrences). The concatenation process of the one-hot encoded text and vectorized categories of nearby location is shown in Fig. 2.

This study is funded by Telekom Malaysia Research & Development

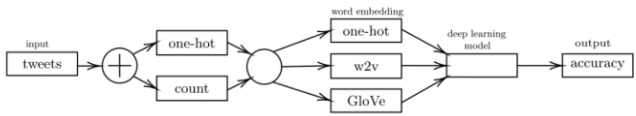

Fig. 2. The concatenation of one-hot encoded text and vectorized categories of nearby location is used as an input to the embedding layer

### E. CNN architecture

In this study, a convolutional neural network (CNN) was used to perform sentiment classification using the concatenated text and geo-location features. The CNN model is composed of one embedding layer, 3 convolution layers followed by max-pooling layer, 1 flatten layer, and 2 dense layers with dropout. Each tweet is represented either as a one-hot encoded vector, or is a concatenation of the one-hot encoded text vector with vectorized geo-location features. For the embedding layer, three pre-trained word embedding models were used, namely word2vec which was trained on Google News, Glove 6B which was trained on Wikipedia, and Glove 27B which was trained on Twitter. Post embedding, three convolution layers with max pooling will generate the internal feature representation before being flattened into a one-dimensional vector. The resultant one-dimensional vector will be passed to two dense layers with dropout with the final layer utilizing a sigmoid activation function to perform the classification. Fig.3 shows the CNN architecture which was used in this study.

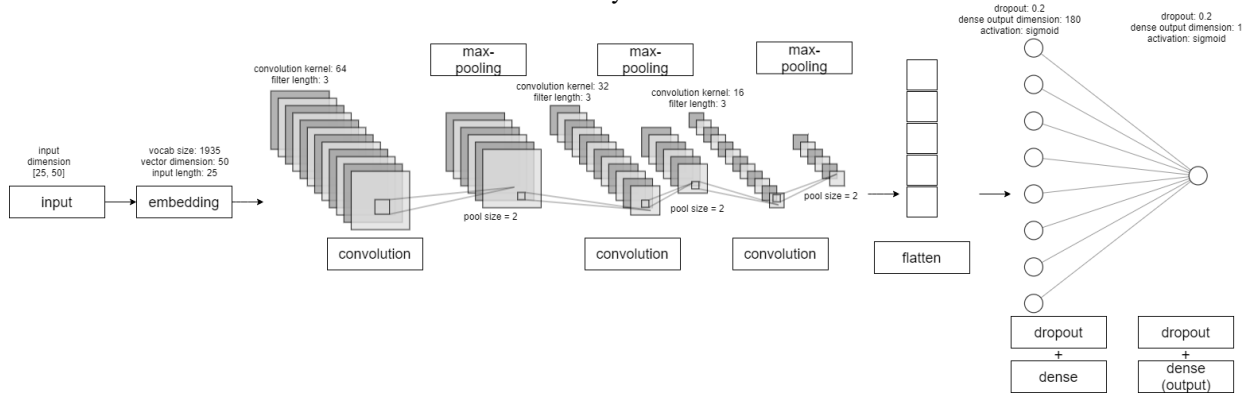

Fig. 3. The CNN architecture for performing tweet sentiment classification using a feature vector of a 50-length concatenated one-hot encoded text and vectorized categories of nearby locations

### F. LSTM architecture

In this study, a long short-term memory (LSTM) was used to perform sentiment classification using the concatenated text and geo-location features. The LSTM model is composed of one embedding layer, one bidirectional LSTM layer, and one dense layer with output. In this model, a bidirectional LSTMs train two , and not just one LSTMs on the concatenated feature post embedding. The first on the input sequence as-is and the second on a reversed copy of the input sequence. For the embedding layer, three pre-trained word embedding models were used, namely word2vec which was trained on Google News, Glove 6B which was trained on Wikipedia, and Glove 27B which was trained on Twitter. Fig.4 shows the LSTM architecture which was in this study.

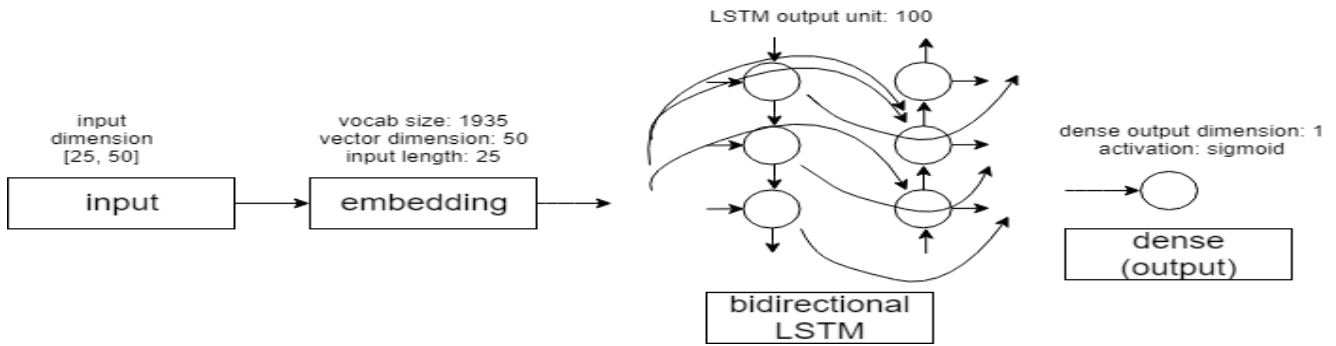

Fig. 4. The CNN architecture for tweet sentiment classification using a feature vector of a 50-length concatenated one-hot encoded text and vectorized categories of nearby locations

## III. EXPRIMENTAL RESULT

The proposed approach of concatenating one-hot encoded text to vectorized categories of nearby location was evaluated using a subset of the 10000 tweets from the Geo-tagged Microblog Corpus. From this subset, 500 geo-tagged tweets with sentiment labels were selected randomly, consisting of 250 positive and 250 negative tweets. This set of 500 labelled tweets were then split into a testing and training set using a 70:30 training to testing split.

The experimental results are grouped according to the deep learning architecture used. The loss (binary cross entropy) and activation function (Adam) is same for both CNN and LSTM experiments. Each experiment is repeat 10 times with 20 epochs.

### A. Dataset

Five variants of dataset were used in this study : (i) tweets only (ii) tweets concatenated with one-hot Geonames nearby location categories (iii) tweets concatenated with one-hot Google Places nearby location categories (iv) tweets concatenated with count of Geonames nearby location categories (v) tweets concatenated with count of Google Places nearby location categories.

### B. Results using CNN

Tables I-IV shows the results of experiments performed, with the best performing experiments for using solely text features underlined, while the best performing experiments using the concatenated text and location features are bolded. In Table 1, the results are shown for experiments conducted with randomized weights for the embedding layer, while Tables II-IV shows the results for experiments performed using weights from the pre-trained embedding models.

TABLE I. ACCURACY OF CNN MODEL USING ONE-HOT EMBEDDING

| Embedding vector dimension | Text only feature | Text & One-hot Geonames features | Text & One-hot Google Places features | Text & Count of Geonames features | Text & Count of Google Places features |
|---|---|---|---|---|---|
| 200 | 69.93 | 70.93 | 66.39 | 68.33 | **71.66** |
| 300 | 69.26 | 69.33 | 66.66 | 69.33 | 70.93 |

TABLE II. ACCURACY OF CNN MODEL USING PRE-TRAINED WORD2VEC WIKIPEDIA 100B EMBEDDING

| Embedding vector dimension | Text only feature | Text & One-hot Geonames features | Text & One-hot Google Places features | Text & Count of Geonames features | Text & Count of Google Places features |
|---|---|---|---|---|---|
| 300 | 71.26 | 73.20 | 73.33 | **73.60** | 72.26 |

TABLE III. ACCURACY OF CNN MODEL USING PRE-TRAINED GLOVE TWITTER 6B EMBEDDING

| Embedding vector dimension | Text only feature | Text & One-hot Geonames features | Text & One-hot Google Places features | Text & Count of Geonames features | Text & Count of Google Places features |
|---|---|---|---|---|---|
| 200 | 69.86 | 70.00 | 69.99 | 70.93 | 69.46 |
| 300 | 70.60 | 71.79 | **72.00** | 71.26 | 70.60 |

TABLE IV. ACCURACY OF CNN MODEL PRE-TRAINED GLOVE TWITTER 27B EMBEDDING

| Embedding vector dimension | Text only feature | Text & One-hot Geonames features | Text & One-hot Google Places features | Text & Count of Geonames features | Text & Count of Google Places features |
|---|---|---|---|---|---|
| 200 | 76.40 | **76.46** | 76.40 | 75.96 | 75.46 |

### C. Results using LSTM

Tables V-VIII shows the results of experiments performed, with the best performing experiments for using solely text features underlined, while the best performing experiments using the concatenated text and location features are bolded. In Table V, the results are shown for experiments conducted with randomized weights for the embedding layer, while Tables VI-VIII shows the results for experiments performed using weights from the pre-trained embedding models.

TABLE V. ACCURACY OF CNN MODEL USING ONE-HOT EMBEDDING

| Embedding vector dimension | Text only feature | Text & One-hot Geonames features | Text & One-hot Google Places features | Text & Count of Geonames features | Text & Count of Google Places features |
|---|---|---|---|---|---|
| 200 | 69.93 | 70.93 | 66.39 | 68.33 | **71.66** |
| 300 | 69.26 | 69.33 | 66.66 | 69.33 | 70.93 |

TABLE VI. ACCURACY OF LSTM MODEL USING PRE-TRAINED WORD2VEC WIKIPEDIA 100B EMBEDDING

| Embedding vector dimension | Text only feature | Text & One-hot Geonames features | Text & One-hot Google Places features | Text & Count of Geonames features | Text & Count of Google Places features |
|---|---|---|---|---|---|
| 300 | 72.60 | 74.20 | 74.53 | 74.53 | **74.86** |

TABLE VII. ACCURACY OF LSTM MODEL USING PRE-TRAINED GLOVE TWITTER 6B EMBEDDING

| Embedding vector dimension | Text only feature | Text & One-hot Geonames features | Text & One-hot Google Places features | Text & Count of Geonames features | Text & Count of Google Places features |
|---|---|---|---|---|---|
| 200 | 76.73 | 71.26 | 71.53 | 70.86 | 70.26 |
| 300 | 79.26 | **74.33** | 73.86 | 74.26 | 73.93 |

TABLE VIII. ACCURACY OF LSTM MODEL PRE-TRAINED GLOVE TWITTER 27B EMBEDDING

| Embedding vector dimension | Text only feature | Text & One-hot Geonames features | Text & One-hot Google Places features | Text & Count of Geonames features | Text & Count of Google Places features |
|---|---|---|---|---|---|
| 200 | 76.93 | 73.59 | 74.46 | **77.99** | 73.06 |

## IV. DISCUSSION

The experiments performed have shown that the proposed approach of using concatenated feature vector consisting of one-hot integer encoded tweets and vectorized nearby categories of nearby locations have resulted in an increase in the accuracy of tweet sentiment classification.

The CNN experimental results showed that on average, the proposed concatenated feature vector consisting of one-hot integer encoded tweets and vectorized nearby categories of nearby locations is higher than that of using purely text features for all concatenated vectors used, except for the concatenation of one-hot text and one-hot Google Places nearby locations. The addition of geo-location information has enriched the initial one-hot vector representation – allowing for the convolution filters to detect additional useful local features.

The LSTM experimental results showed the opposite results, with the pure text features on average outperforming the proposed concatenated feature vector. This outcome is heavily impacted by model which used the pre-trained Glove 6B model which fitted badly to the tweet data used. This outcome suggests that sequential information is of less importance to the geo-tagged tweet sentiment classification task.

In terms of API performances, it seems that there is no differences in using either the Google Places or Geonames web-services , as using both APIs gave the best results in 50% of the architectures used in the experiments.

For the vectorization approach for the category of the nearby locations, the count method yielded better results than the one-hot approach, with 5 out of 8 deep learning architectures used showing better results using the count approach.

We also observed that the size of the pre-trained word embedding models played with the 100B and 27B word embedding models outperforming the 6B word embedding model.

## V. CONCLUSION

In this study, we have proposed a new approach towards classification of geo-tagged tweets, using a concatenated feature vector consisting of one-hot integer encoded tweets and vectorized nearby categories of nearby locations. A significant challenge is in the interpretation of the effects of adding geo-location information to the text representation.

The addition of location features has resulted in an increase in accuracy rates in the experiments performed. For the experiments involving CNN, a total of 5.53% improvement was achieved when the concatenated word and location features were used as opposed to solely using one-hot encoded text vectors. RNN showed a similar net improvement when all experiments were considered with an improvement of 0.12% percent.

For future works, we will attempt to use a larger dataset with mixed language content, as well as develop a mechanism which will help to better explain and interpret the models developed using our approach.